\documentclass{article}
\usepackage{spconf,amsmath,graphicx}
\usepackage{amsmath,amssymb,amsfonts}
\usepackage{multirow}
\usepackage{algorithm}
\usepackage[algo2e]{algorithm2e}
\usepackage{algorithm}
\usepackage{algpseudocode}
\usepackage{xcolor}
\title{Moving Object Classification with a Sub-6 GHz Massive MIMO Array using Real Data \vspace*{-0.13in}}
\name{B. R. Manoj$^{\dagger}$, Guoda Tian$^{\ast}$, Sara Gunnarsson$^{\ast}$, Fredrik Tufvesson$^{\ast}$, Erik G. Larsson$^{\dagger}$
\vspace*{-0.1in}
\thanks{This work was supported in part by ELLIIT, Security-Link and a grant from Ericsson AB.}}
\address{$^{\dagger}$Department of Electrical Engineering (ISY), Link{\"o}ping University, Link{\"o}ping, Sweden\\
$^{\ast}$Department of Electrical and Information Technology, Lund University, Lund, Sweden \\
\small{Emails: $\text{\{manoj.banugondi.rajashekara, erik.g.larsson\}@liu.se}$, $\text{\{guoda.tian, sara.gunnarsson, fredrik.tufvesson\}}\text{@eit.lth.se}$} %$\text{erik.g.larsson@liu.se}$
\vspace*{-0.25in}}
\begin{document}
\maketitle
\begin{abstract}
\vspace*{-0.05in}
Classification between different activities in an indoor environment using wireless signals is an emerging technology for various applications, including intrusion detection, patient care, and smart home. Researchers have shown different methods to classify activities and their potential benefits by utilizing WiFi signals. In this paper, we analyze classification of moving objects by employing machine learning on real data from a massive multi-input-multi-output (MIMO) system in an indoor environment. We conduct measurements for different activities in both line-of-sight and non line-of-sight scenarios with a massive MIMO testbed operating at 3.7 GHz. We propose algorithms to exploit amplitude and phase-based features 
classification task. For the considered setup, we benchmark the classification performance and show that we can achieve up to 98\% accuracy using real massive MIMO data, even with a small number of experiments. Furthermore, we demonstrate the gain in performance results with a massive MIMO system as compared with that of a limited number of antennas such as in WiFi devices. 
\end{abstract}
\vspace*{-0.05in}
\vspace*{-0.17in}
\section{Introduction}
\label{sec:intro}
\vspace*{-0.11in}
Wireless-based activity sensing technology is evolving rapidly due to the interest in many day-to-day applications such as intrusion detection, patient care, etc., \cite{Liu_wireless_2020} without requiring %using devices such as 
cameras, motion sensors, or radars \cite{Wang_device_2017}. The potential of utilizing wireless signals for activity sensing has been shown for a variety of applications such as non line-of-sight (NLOS)/line-of-sight (LOS) identification \cite{Wu_phaseu_2015,Huang_machine_2020}, motion detection \cite{Xiao_fimd_2012}, human presence detection \cite{Zhu_RTTWD_2017,Zhang_wifi-id_2016}, classification of human activities \cite{Wang_device_2017,Ding_wihi_2020,Zhang_wivi_2020} and localization \cite{Kotaru_spotfi_2015}.  These can be realized indoors by measuring channel state information (CSI) from already available WiFi devices \cite{Xiao_fimd_2012}--\cite{Zhang_wivi_2020}. 
The machine learning (ML)-based approaches have shown success and been widely used in numerous applications due to their ability to learn the statistical patterns from the channel information. A few examples of applications are 
%The performance of activities classification using WiFi CSI can be investigated by learning the statistical patterns through machine learning (ML) models. The ML approaches have shown success in numerous applications, a few examples being 
human activities recognition using a hidden Markov model \cite{Wang_device_2017}, NLOS/LOS identification for multiple-input-multiple-output (MIMO) systems \cite{Wu_phaseu_2015,Huang_machine_2020}, massive MIMO-based indoor positioning \cite{Arnold_novel_2019, Bast_csi_2020}, 
and radio identification \cite{Riyaz_deep_2018}. Although ML approaches have shown success in WiFi-based activity sensing as well, the performance is limited since the devices typically are equipped with two or three antennas. This limitation imposes difficulties in exploiting statistical patterns in the spatial domain and degraded accuracy especially in NLOS conditions.  
%due to sensitivity of the limited number of links. 

In this paper, we aim to overcome these limitations by using real data from a massive MIMO system to exploit the advantages of the spatial domain to efficiently classify moving objects in LOS and NLOS scenarios by using ML models. To the best of our knowledge, this has not yet been addressed in the literature. To exploit the multiple antennas, we utilize the information from correlation changes across space as the feature for activity detection. The main contributions of this paper are: 
(i) We propose algorithms to extract features from the amplitude and phase information of the measured massive MIMO data. The measurements include LOS and NLOS propagation scenarios for both static and dynamic environments. In the literature, it is often only the amplitude information that is considered. A few methods have been proposed to also exploit phase information, such as linear transformation and phase difference between the antennas \cite{Wu_phaseu_2015}. Different from existing techniques, we consider relative phase changes by applying a linear regression method of phase trajectories to obtain the error variance and then using its correlation changes across space as the phase-based feature. (ii) We compare two different ML methods for classifying different moving objects, namely support vector machine (SVM) and feedforward neural network (NN) with fully connected layers.
\vspace*{-0.26in}
\section{Moving Object Classification}
\label{sec:algo}
\vspace*{-0.11in}
We consider an uplink narrow band massive MIMO orthogonal frequency division multiplexing (OFDM) system with  multiple  user equipments (UEs).
%, where each single antenna UE allocates different subcarriers. 
The received signal matrix $\mathbf{Y}_f\in \mathbb{C}^{M\times N}$ for all antennas $M$ and snapshots $N$ can, for each subcarrier $f$, be described as
\vspace{-5pt}
\begin{equation}
    \label{receive}
    \mathbf{Y}_f = \mathbf{H}_{f} \odot\mathbf{\Gamma}_f\hspace{1pt} + \mathbf{N}_f\, ,
    \vspace{-5pt}
\end{equation}
where subcarrier $f \in [1,F]$, radio-frequency (RF)-chain $m \in [1, M]$, snapshot $n\in [1, N]$, $\odot$ is the Hadamard product and $\mathbf{H}_{f}\in \mathbb{C}^{M\times N}$ is the complex-valued channel matrix. It is difficult to precisely model $\mathbf{H}_{f}$ due to unknown positions of the UEs and the environment as well as unknown Doppler shifts caused by unpredictable speeds and directions of different moving objects.
Furthermore, we denote the frequency response of the RF chains as $\mathbf{\Gamma}_f\in \mathbb{C}^{M\times N}$ with each element defined as $\mathbf{\Gamma}_f(m,n) = d_m e^{j(\alpha_m-n\,\epsilon_{m,f})}$, where $d_m$, $\alpha_m$, $\epsilon_{m,f}$ represent the amplitude scaling, initial phase offset, and carrier-frequency-offset (CFO) for the $f$-th subcarrier, respectively, for the $m$-th RF chain. 
%The CFO varies for each subcarrier due to the imperfect clock synchronization between the UE and base station (BS). 
The noise for the $f$-th subcarrier is denoted as $\mathbf{N}_f\in \mathbb{C}^{M\times N}$. Finally, for a total of $F$ subcarriers, we define $\mathcal{Y} \in \mathbb{C}^{F\times M\times N}$ to represent the received data. %Since we do not have a physically accurate signal model, it is challenging to
Since we have limitations in the signal model, it is challenging to apply traditional detection and estimation theory for the   
considered classification task. Thus, we are motivated to exploit the statistical features that can be efficiently used by  ML models to accurately classify the moving objects.
\vspace*{-0.3in}
\subsection{Feature extraction}
\label{ssec:feature_extraction}
\vspace*{-0.08in}
For many wireless-based applications, the ML models can operate with raw I/Q samples. However, this requires a huge data set for training the models, which in turn increases hardware requirements and computational complexity. Instead of using the raw I/Q samples as input, we exploit features from a real data set to reduce the dimensionality, which significantly reduces the training data set.
In this section, we propose algorithms to extract these features from the amplitude and phase information. Before this,  
the first step is to apply linear interpolation on the data to overcome potential problems of sampling jitter during measurements \cite{Zhu_RTTWD_2017}, i.e., to obtain evenly-spaced samples. 
\vspace*{-0.15in}
\subsubsection{Amplitude-based feature}
\vspace*{-0.08in}
After the interpolation step, to eliminate random noise present in the data, we use the discrete wavelet-based denoising method \cite{Zhu_RTTWD_2017,Ding_wihi_2020}. 
Similar to \cite{Zhu_RTTWD_2017}, we use a 2-level  
wavelet transform on the amplitude of the data of all subcarriers across the snapshots for each antenna. We then apply the principle of Stein's unbiased risk estimate thresholding to the high-frequency coefficients to filter out the noisy part \cite{Zhu_RTTWD_2017}. We denote the $3$-dimensional data after processing with noise filtering as $|{\cal{Z}}| \in {\mathbb{R}}^{F \times M \times N}$.
In the literature, researchers have utilized various statistical features in the time and frequency domain for activity detection \cite{Zhang_wifi-id_2016, Ding_wihi_2020}. Different from previous works, we exploit correlation changes not only across time and frequency, but also space, which occur during the object movement. We further utilize principal component analysis (PCA) to track these changes. For this, we choose a time window ($T_w$) of $1$~second, to capture the small movement changes of the object in the area of interest, such that we accurately can estimate the time-varying correlation across the frequency and space in $T_w$. Algorithm~$1$, presents our proposed method for amplitude-based feature extraction. In the algorithm, we discard the first eigenvalue since it has been observed that its value is high for both static and dynamic events, making it difficult to differentiate them. \vspace{-16pt}
\begin{algorithm}[t]
\scriptsize
 	\label{alg_amplitude}
	\SetAlgoLined
	\textbf{Input:} $|{\cal{Z}}| \in {\mathbb{R}}^{F \times M \times N}$ and $T_w$, \quad \textbf{Output:} Amplitude feature, ${\mathcal{A}}$
	
	\For {$n=1$ $\mathrm{to}$ $N$}
	{
	Define $\mathbf{B}_n \in {\mathbb{R}}^{F \times M}$, as the matrix obtained from $|{\cal{Z}}|$ 
	at time $n$:\\
	$\mathbf{B}_n = [\mathbf{b}_n(1), \hdots, \mathbf{b}_n(F)]^\mathrm{T}$, $\mathbf{b}_n(f) \in {\mathbb{R}}^{M}$, $f \in [1,F]$\,.
	}
	Define $\mathbf{D}\in {\mathbb{R}}^{FM \times N}$, as the matrix obtained from vectorizing the matrices $\mathbf{B}_n$, $n \in [1,N]$: \\
	$\mathbf{D} = [\mathrm{vec}(\mathbf{B}_1), \mathrm{vec}(\mathbf{B}_2),\hdots, \mathrm{vec}(\mathbf{B}_N)]$\,.
	
	\For {$j=1$ $\mathrm{to}$ $N/T_w$}
	{
	 $\mathbf{E} = [\mathbf{D}(1), \mathbf{D}(2), \hdots, \mathbf{D}(T_w)]$, $\mathbf{E} \in {\mathbb{R}}^{FM \times T_w}, \mathbf{D}(i)\in {\mathbb{R}}^{FM \times 1}, i \in [1, T_w]$\,.\\
	 Determine the inner product: 
     $\mathbf{S} = \mathbf{E}^\mathrm{T} \mathbf{E}$, $\mathbf{S} \in {\mathbb{R}}^{T_w \times T_w}$\,. \\
     Perform eigenvalue decomposition: $\mathbf{S} = \mathbf{U} \mathbf{\Sigma} \mathbf{U}^\mathrm{T}$\,.\\ 
     $\mathbf{g}_j$ = Sort the eigenvalues in the descending order. Discard the first eigenvalue and store the rest. \\
     Slide $T_w$ in $\mathbf{D}$ and repeat the calculations of $\mathbf{S}$ and $\mathbf{\Sigma}$.
    }
	$\mathbf{G} =[\mathbf{g}_1, \mathbf{g}_2, \hdots, \mathbf{g}_{N/T_w}]^\mathrm{T}$\,. %\\
	\,\, ${\mathcal{A}} = \mathbb{E}[\mathbf{G}]$, where $\mathbb{E}[\cdot]$ is the expectation operator.
	\caption{Amplitude-based feature extraction}
\end{algorithm}
\vspace*{0.1in}
\subsubsection{Phase-based feature}
\vspace*{-0.08in}
\begin{figure}[t]
\vspace*{-0.15in}
  \centering
  \centerline{\includegraphics[scale=0.58]{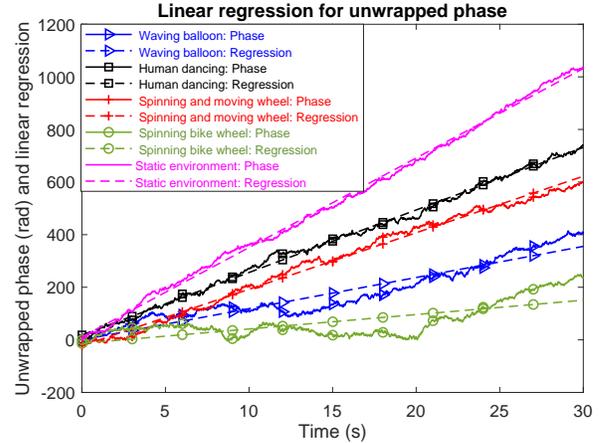}\vspace*{-0.18in}}
  \caption{Unwrapped phase and linear regression for $f=30$ and $m=1$.\vspace*{-0.2in}}
  \label{fig:phase_reg}
\end{figure}
The phase information is more sensitive for different types of moving objects, thus, playing an important role to improve classification accuracy. However, for a static scenario, only minor phase changes are expected due to measurement noise. For dynamic events, the phase of $\mathbf{H}_{f}$ 
changes more rapidly across the snapshots due to Doppler shifts. 
Furthermore, the phase of $\mathbf{\Gamma}_f(m,n)$ increases or decreases  linearly with snapshots due to the CFO across the subcarriers. This phase component contributes to the raw unwrapped phase of $\mathcal{Y}$, which varies around the linear regression line.
We, therefore, first perform linear regression on the unwrapped phase of $\mathcal{Y}$ for each $f$ and 
$m$  across all snapshots. 
The unwrapped phase and the linear regression w.r.t. the observation time of different events for $f = 30$ and $m = 1$ are demonstrated in Fig. \ref{fig:phase_reg}. Since the performance of the CFO and timing estimation varies, the CFOs, after compensation between the UE and BS, could be different between each sample, resulting in non-identical slopes as shown in Fig. \ref{fig:phase_reg}. 
We denote $\hat{\mathcal{Y}}$ as the unwrapped phase of $\mathcal{Y}$ and $\hat{\mathbf{y}}_{f,m}\in \mathbb{R}^N$ as the vector representing the phase snapshot obtained from $\hat{\mathcal{Y}}$ for the $f$-th subcarrier and $m$-th RF chain. We define $\boldsymbol{\xi} = [1,2,...,N]^\mathrm{T} \in \mathbb{R}^{N}$ as the indexes of the snapshots, $\mathbf{1}_N\in \mathbb{R}^{N}$ as the column unit vector, and $\mathbf{\Xi}\in \mathbb{R}^{N \times 2}$ as $\mathbf{\Xi} = \left[\mathbf{1}_N, \hspace{2pt} \boldsymbol{\xi}\right]$. Algorithm~$2$ presents our proposed method for the phase-based feature extraction. In the algorithm,  $\boldsymbol{\beta}_{f,m} \in \mathbb{R}^{2}$ is the linear regression vector, obtained using least squares method \cite{Inghelbrecht_large_2020}, of $\hat{\mathbf{y}}_{f,m}$. 
\begin{algorithm}[t]
\scriptsize
 	\label{alg_phase}
	\SetAlgoLined
	\textbf{Input:} Unwrapped phase $\hat{\mathcal{Y}}\in {\mathbb{R}}^{F \times M \times N}$, \quad 	\textbf{Output:} Phase feature, ${\mathcal{P}}$\\
	\For {$m=1$ $\mathrm{to}$ $M$}
	{
      \For {$f=1$ $\mathrm{to}$ $F$}
	 {
	   Linear regression for each $f$ and $m$: $\boldsymbol{\beta}_{f,m} = (\mathbf{\Xi}^\mathrm{T}\hspace{1pt}\mathbf{\Xi})^{-1}\hspace{1pt}\mathbf{\Xi}^\mathrm{T}\hspace{1pt}\hat{\mathbf{y}}_{f,m}$. \\
	   Define $\boldsymbol{\eta}_{f,m}\in \mathbb{R}^N$ as the deviation between $\hat{\boldsymbol{y}}_{f,m}$ and the regression line: \\
	   ${\boldsymbol{\eta}}_{f,m}=\hat{\boldsymbol{y}}_{f,m} -  \boldsymbol{\beta}_{f,m}(2) \cdot \boldsymbol{\xi} - \boldsymbol{\beta}_{f,m}(1) \cdot \mathbf{1}_N$\\
	   Define $q_{f,m}$ as the variance of ${\boldsymbol{\eta}}_{f,m}$: $q_{f,m} = \mathrm{var}({\boldsymbol{\eta}}_{f,m})$
	}}
	Define $\mathbf{Q}\in \mathbb{R}^{F\times M}$, where the $f$-th row and $m$-th column of $\mathbf{Q}$ is $q_{f,m}$.\\
	Calculate the pairwise column correlation of $\mathbf{Q}$: $\tilde{\mathbf{S}} \in \mathbb{R}^{M \times M}$.\\ 
    Perform eigenvalue decomposition: $\tilde{\mathbf{S}} = \tilde{\mathbf{U}} \tilde{\mathbf{\Sigma}} \tilde{\mathbf{U}}^\mathrm{T}$. \\
    Sort the eigenvalues in the descending order and discard the first eigenvalue. The rest are stored in ${\mathcal{P}}$ as the phase-based features. 
 \caption{Phased-based feature extraction}
 %\vspace*{-0.1in}
 \end{algorithm} 
\vspace*{-0.18in}
\subsection{Machine learning models}
\label{ssec:ml_models}
\vspace*{-0.05in}
%In this section, 
We describe the ML models SVM and feedforward NN that is applied for the classification task by leveraging the features extracted from the proposed algorithms. We denote the classifier model as $f(\mathbf{x}): \mathbf{x} \in \mathcal{X} \rightarrow \mathbf{c} \in \mathcal{C}$, where the input ${\mathbf{x}} = [\mathcal{A}, \mathcal{P}]$ is a combination of the aforementioned amplitude and phase-based features while $\mathbf{c}$ is the predicted output label. The training data set is defined as $\{\mathbf{x}(t),\mathbf{c}(t)\}_{t=1}^{N_T}$, where $N_T$ is the size of data set; the ML models are trained with this data set to learn the relation between the input features and the output label. The details of the considered ML models are as follows (i) SVM: This is a classical supervised ML model used for classification tasks. It is widely used since it requires a few predefined parameters and uses a simple optimization problem to learn the weights of the model. We implement our algorithm using the SVM model by utilizing the package $sklearn$ \cite{scikit-learn} with the kernel type $linear$. Since it is difficult to design the kernel type that models the nonlinearity in the data set, we resort to the standard linear kernel. (ii) Feedforward NN: In comparison to SVM, the NN transforms the input through linear mapping and then by using the nonlinear activation function. Due to this nature of NN, it makes it efficient to learn the nonlinearity in the data set compared to classical methods. We consider a simple feedforward architecture with fully connected layers consisting of an input layer, a few hidden layers, and an output layer. The detailed description of NN 
architecture that is proposed is given in Table \ref{tab_nn}. For this architecture, we have considered the Adam optimizer and the loss function as categorical cross-entropy, and model parameters have been designed such that the network performs efficiently. The maximum input dimension to the SVM and NN architectures, shown in Table \ref{tab_nn}, is set to $12$, as the second to seventh largest
eigenvalues of the amplitude- and phase-based features. Through our experiments, we have observed that a further increase in the input dimension does not improve the learning ability of the ML models. 
\begin{table}[t!]
  \begin{center}
  \vspace{-0.5em}
    \caption{NN architecture with trainable parameters of $3,576$.}
    \label{tab_nn} \scalebox{0.64}{
    \begin{tabular}{|c|c|c|c|}  
        & Size & Parameters & Activation function \\
      \hline
      Input: [${\mathcal{A}},\,{\mathcal{P}}$] & 12 & - & - \\  \hline
      Layer 1 (Dense) & 64 & 832  & elu \\ \hline
      Layer 2 (Dense) & 32 & 2080 & elu \\ \hline
      Layer 3 (Dense) & 16 & 528  & elu \\ \hline
      Layer 4 (Dense) & 8  & 136  & elu \\ \hline
      Layer 5 (Dense) & 2  & 18  & softmax \\ \hline
      \end{tabular} }
  \end{center}
  \vspace*{-0.28in}
\end{table}
\vspace*{-0.15in}
\section{Massive MIMO measurement setup}
\label{sec:measurement}
\begin{figure}[t]
  \centering
  \centerline{\includegraphics[scale=0.4]{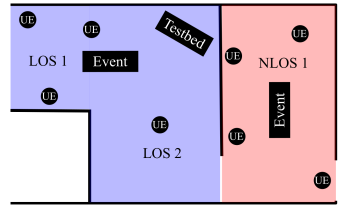}\vspace*{-0.15in}}
  \caption{Map over the scenarios. For the LOS scenario, the UEs are placed in LOS~$1$ and~$2$; for the NLOS scenario, they are placed in NLOS~$1$. In case of a dynamic environment, the activity takes place in the ``Event" box. \vspace*{-0.1in}}
  \vspace*{-0.1in}
  \label{fig:los_nlos_map}
\end{figure}

\vspace*{-0.13in}
A measurement campaign has been carried out in an indoor laboratory environment at Lund University, to capture samples under both LOS and NLOS scenarios. The measurement setup is illustrated in Fig. \ref{fig:los_nlos_map}, where the UEs and the BS are static and placed in the same room in the LOS scenario, while separated by a wall in the NLOS scenario.
For each of the scenarios, four different dynamic events were performed, namely, waving an aluminium foil balloon, spinning a bike wheel, spinning and moving a bike wheel, and human dancing. Samples were also collected in static environments. In a given scenario, for each of the static and dynamic events, 18~experiments were conducted, resulting in a total of 90~experiments. 

The BS is the Lund University massive MIMO testbed (LuMaMi) \cite{lumami_journal}, which is a software-defined radio-based testbed that operates in OFDM-mode with $100$~antennas connected to $100$ transceiver chains at a carrier frequency of $3.7$~GHz with $20$~MHz of bandwidth. The antenna elements are separated half a wavelength apart and arranged in four rows of $25$ elements each; counting row-wise the odd-numbered elements are vertically polarized and the even-numbered are horizontally polarized. 
The UEs consist of an universal software radio peripheral (USRP) with two transceiver chains and are equipped with either one or two dipole antennas. For each measurement and active UE transceiver chain, $100$~frequency points, and $3000$~snapshots over $30$~seconds were collected, this constituting one experiment.
\vspace*{-0.27in}
\section{Results and discussion}
\label{ec:experiments_results}
\begin{figure}[!t]
\begin{minipage}[b]{0.49\linewidth}
  \centering
  \centerline{\includegraphics[width=1.05\linewidth]{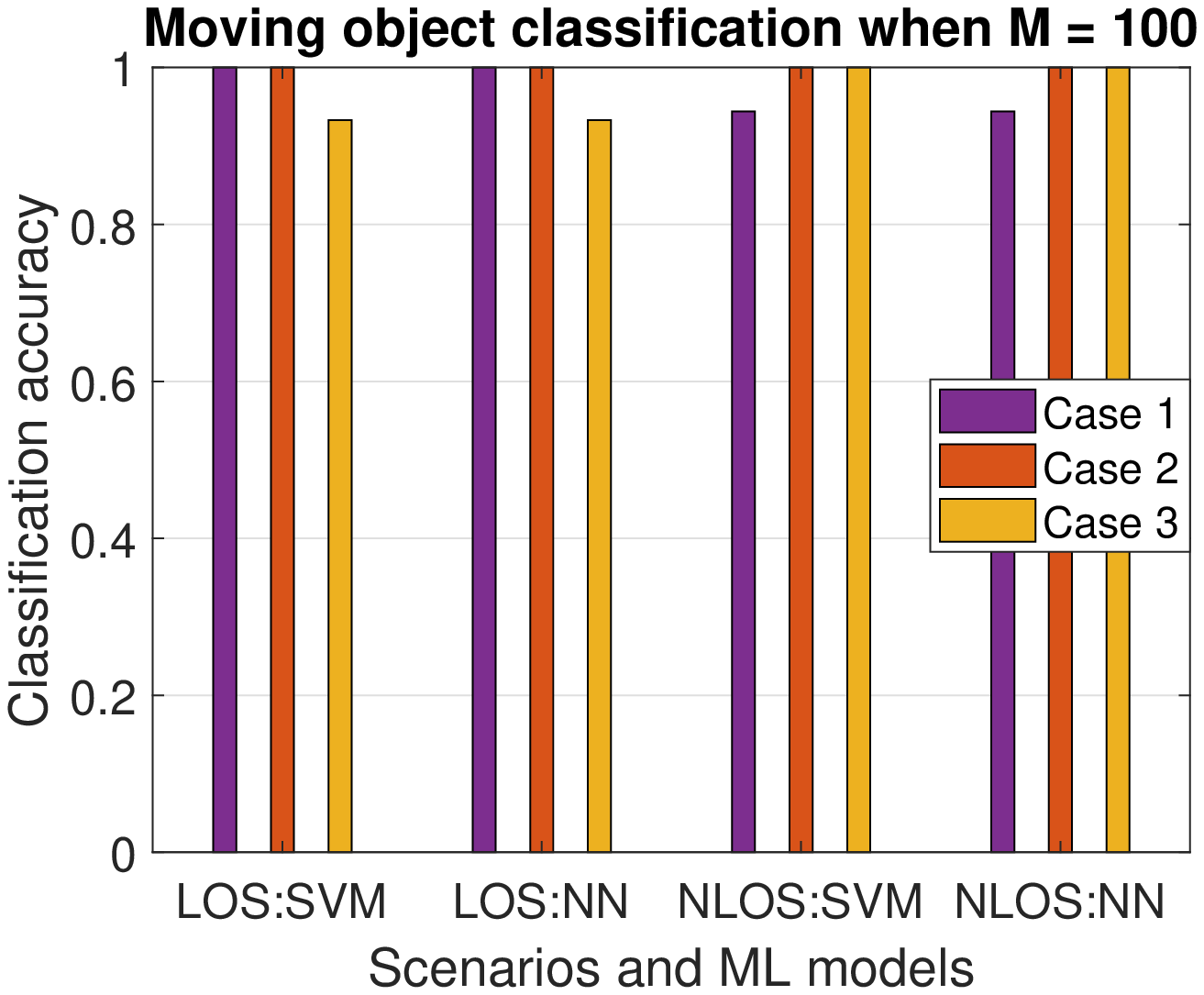}}
\vspace{-.1em}
  \centerline{(a)}\medskip
\end{minipage}
\begin{minipage}[b]{0.49\linewidth}
  \centering
  \centerline{\includegraphics[width=1.05\linewidth]{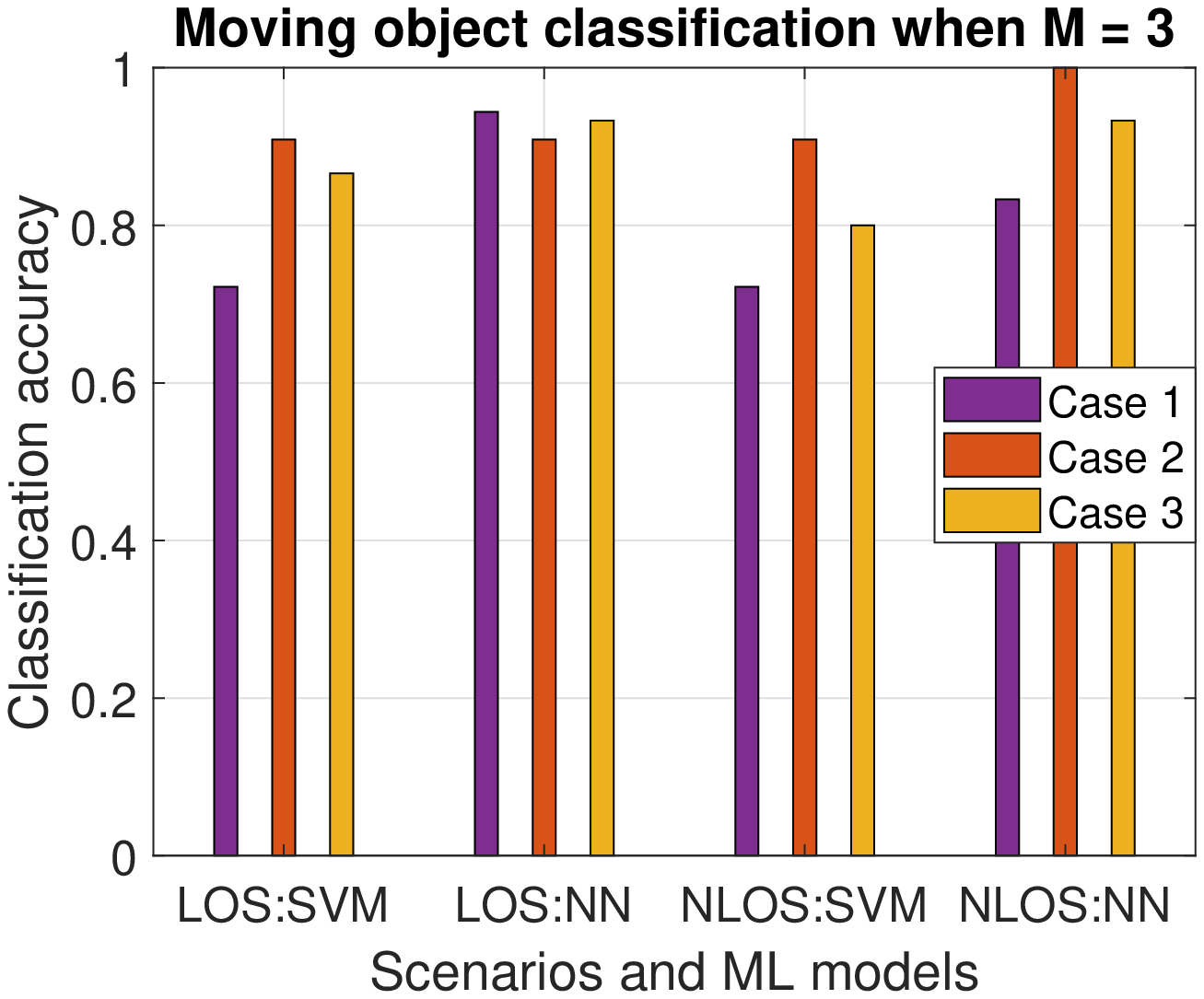}}
\vspace{-0.1em}
  \centerline{(b)}\medskip
\end{minipage}
\vspace{-1.5em}
\caption{Classification accuracy of Case 1--3 in LOS and NLOS scenarios using SVM and NN models for (a) $M = 100$ and (b) $M = 3$.}
\label{fig:classification_acc}
\vspace*{-0.2in}
\end{figure}

\vspace*{-0.13in}
In this section, we evaluate the %moving objects 
classification performance using the proposed algorithms for feature extraction from the  massive MIMO measurements. We denote the measurement events as
$v_1$:~static, $v_2$:~human dancing, $v_3$:~spinning bike wheel, $v_4$:~waving of an aluminium foil balloon, and $v_5$:~spinning and moving bike wheel.  
We consider the classification problems as depicted in~Table \ref{tab_classification}. The extracted features from the events are labelled as shown in the third column of Table \ref{tab_classification}. In this table, Case 1 depicts the scenario of classifying between static and dynamic environments, Case 2 between a human dancing activity and one non-human moving object and Case 3 is similar to Case 2, besides, we consider all non-human moving objects. 
\begin{table}[t]
  \begin{center}
  \vspace{-0.25em}
    \caption{Classification of different moving objects.}
    \label{tab_classification} \scalebox{0.65}{
    \begin{tabular}{|c|c|c|}  
    \hline 
         Cases & Classifications & Labels \\
      \hline
1 & Between $\{v_2, v_3, v_4, v_5\}$ and $\{v_1\}$ & $\{v_2, v_3, v_4, v_5\} \rightarrow$ `1' \\ 
  & & $\{v_1\} \rightarrow$ `0' \\ \hline 
2 & Between $\{v_2\}$ and $\{v_3\}$ & $\{v_2\} \rightarrow$ `1'  \\ 
& & $\{v_3\} \rightarrow$ `0' \\ \hline 
3 & Between $\{v_2\}$ and $\{v_3, v_4, v_5\}$ & $\{v_2\} \rightarrow$ `1' \\
& & $\{v_3, v_4, v_5\} \rightarrow$ `0' \\ \hline 
      \end{tabular} }
  \end{center}
  \vspace*{-0.33in}
\end{table}
The classification accuracy when the number of antennas $M = 100$ and $M = 3$ for Case 1--3 are shown in Figs. \ref{fig:classification_acc}(a) and \ref{fig:classification_acc}(b), respectively.
For training and testing (train:test) the ML models, we consider (80\%:20\%), (70\%:30\%), and (80\%:20\%) for Cases 1, 2, and 3, respectively. The reason for choosing different values is due to the different sizes of the data sets.   
%is based on the availability of the current data set. 
The objective of this paper is to showcase the potential of using massive MIMO for moving objects classification; thus, it has to be noted that the results presented here are preliminary as the number of experiments is small, though from Fig. \ref{fig:classification_acc}(a) the results seem to be promising, further investigation with more diverse measurements is needed. 

For Case 1--3, we have presented the confusion matrices in Table~\ref{tab:Case1_M_100}--\ref{tab:Case3_M_100} when $M = 100$. 
In the tables, the diagonal entries represent the number of samples that are correctly classified whereas off-diagonal entries represent the number of misclassified samples. From Fig. \ref{fig:classification_acc}(b), for both LOS and NLOS scenarios, the NN outperforms the SVM based method, since SVM is not able to capture the nonlinearity in the data set for different dynamic activities. The careful selection of hyperparameters and hidden layers for linear mapping and then the nonlinear activation function in NN enables the model to learn the input and output relationship better, thus, NN achieves
good classification accuracy. 
In Table \ref{tab_nn}, we have shown the number of trainable parameters assuming the upper limit on the input dimension, i.e., ${\mathbf{x}} = [6,6]$. We have observed through empirical experiments that for Case 2, we need only ${\mathbf{x}} = [2,2]$. For comparison, the classification accuracy results for $M = 3$ is depicted in Fig. \ref{fig:classification_acc}(b). As expected, it can be observed that the case with $M = 100$ outperforms $M = 3$, since the spatial correlation changes are exploited better with the large antenna array. We have also presented the confusion matrices in Table \ref{tab:Case1_M_3}--\ref{tab:Case3_M_3} when $M = 3$ for Case 1--3. From the confusion matrices, it is evident that the probability of misclassification of the events is higher when $M = 3$ as compared to $M = 100$. 
%%%%%%%%%%%%%%%%%%%%%%%%%%%%%%%%%%%%%%%%%%%%%%%%%%%%%%%%%%
\vspace*{-0.2in}
\begin{table}[t!]
\caption{Confusion matrices for Case 1 when $M = 100$.}
\vspace{2pt}
\label{tab:Case1_M_100}
\center
\scalebox{0.56}{
\begin{tabular}{|l|l|l|l|l|l|l|l|}
\hline
\multirow{3}{*}{\begin{tabular}[c]{@{}l@{}}LOS\\ using\\ SVM\end{tabular}} &
   &
  $\{v_1\}$ &
  $\{v_2, v_3, v_4, v_5\}$ &
  \multirow{3}{*}{\begin{tabular}[c]{@{}l@{}}NLOS\\ using\\ SVM\end{tabular}} &
   &
  $\{v_1\}$ &
  $\{v_2, v_3, v_4, v_5\}$ \\ \cline{2-4} \cline{6-8} 
 & $\{v_1\}$  & 5 &0  &  & $\{v_1\}$  & 5 & 0  \\ \cline{2-4} \cline{6-8} 
 & $\{v_2, v_3, v_4, v_5\}$  & 0 & 13 &  & $\{v_2, v_3, v_4, v_5\}$  & 1 & 12 \\ \hline
\multirow{3}{*}{\begin{tabular}[c]{@{}l@{}}LOS\\ using\\ NN\end{tabular}} &
   &
  $\{v_1\}$ &
  $\{v_2, v_3, v_4, v_5\}$  &
  \multirow{3}{*}{\begin{tabular}[c]{@{}l@{}}NLOS\\ using\\ NN\end{tabular}} &
   &
  $\{v_1\}$ &
  $\{v_2, v_3, v_4, v_5\}$  \\ \cline{2-4} \cline{6-8} 
 & $\{v_1\}$  & 5 & 0  &  & $\{v_1\}$  & 4 & 1  \\ \cline{2-4} \cline{6-8} 
 & $\{v_2, v_3, v_4, v_5\}$  & 0 & 13 &  & $\{v_2, v_3, v_4, v_5\}$  & 0 & 13 \\ \hline
\end{tabular}}
\vspace*{-0.2in}
\end{table}
\begin{table}[t!]
\caption{Confusion matrices for Case 2 when $M = 100$.}
\vspace{2pt}
\label{tab:Case2_M_100}
\center
\scalebox{0.56}{
\begin{tabular}{|l|l|l|l|l|l|l|l|}
\hline
\multirow{3}{*}{\begin{tabular}[c]{@{}l@{}}LOS\\ using\\ SVM\end{tabular}} &
   &
  $\{v_3\}$ &
  $\{v_2\}$ &
  \multirow{3}{*}{\begin{tabular}[c]{@{}l@{}}NLOS\\ using\\ SVM\end{tabular}} &
   &
  $\{v_3\}$ &
  $\{v_2\}$ \\ \cline{2-4} \cline{6-8} 
 & $\{v_3\}$  & 8 & 0  &  & $\{v_3\}$  & 8 & 0  \\ \cline{2-4} \cline{6-8} 
 & $\{v_2\}$  & 0 & 3 &  & $\{v_2\}$  & 0 & 3 \\ \hline
\multirow{3}{*}{\begin{tabular}[c]{@{}l@{}}LOS\\ using\\ NN\end{tabular}} &
   &
  $\{v_3\}$ &
  $\{v_2\}$  &
  \multirow{3}{*}{\begin{tabular}[c]{@{}l@{}}NLOS\\ using\\ NN\end{tabular}} &
   &
  $\{v_3\}$ &
  $\{v_2\}$  \\ \cline{2-4} \cline{6-8} 
 & $\{v_3\}$  & 8 & 0  &  & $\{v_3\}$  & 8 & 0  \\ \cline{2-4} \cline{6-8} 
 & $\{v_2\}$  & 0 & 3 &  & $\{v_2\}$  & 0 & 3 \\ \hline
\end{tabular}}
\vspace*{-0.15in}
\end{table}
\begin{table}[t!]
\caption{Confusion matrices for Case 3 when $M = 100$.}
\vspace{2pt}
\label{tab:Case3_M_100}
\center
\scalebox{0.56}{
\begin{tabular}{|l|l|l|l|l|l|l|l|}
\hline
\multirow{3}{*}{\begin{tabular}[c]{@{}l@{}}LOS\\ using\\ SVM\end{tabular}} &
   &
  $\{v_3, v_4, v_5\}$ &
  $\{v_2\}$ &
  \multirow{3}{*}{\begin{tabular}[c]{@{}l@{}}NLOS\\ using\\ SVM\end{tabular}} &
   &
  $\{v_3, v_4, v_5\}$ &
  $\{v_2\}$ \\ \cline{2-4} \cline{6-8} 
 & $\{v_3, v_4, v_5\}$  & 11 & 0  &  & $\{v_3, v_4, v_5\}$  & 11 & 0  \\ \cline{2-4} \cline{6-8} 
 & $\{v_2\}$  & 1 & 3 &  & $\{v_2\}$  & 0 & 4 \\ \hline
\multirow{3}{*}{\begin{tabular}[c]{@{}l@{}}LOS\\ using\\ NN\end{tabular}} &
   &
  $\{v_3, v_4, v_5\}$ &
  $\{v_2\}$  &
  \multirow{3}{*}{\begin{tabular}[c]{@{}l@{}}NLOS\\ using\\ NN\end{tabular}} &
   &
  $\{v_3, v_4, v_5\}$ &
  $\{v_2\}$  \\ \cline{2-4} \cline{6-8} 
 & $\{v_3, v_4, v_5\}$  & 10 & 1  &  & $\{v_3, v_4, v_5\}$  & 11 & 0  \\ \cline{2-4} \cline{6-8} 
 & $\{v_2\}$  & 0 & 4 &  & $\{v_2\}$  & 0 & 4 \\ \hline
\end{tabular}}
\vspace*{-0.15in}
\end{table}
%%%%%%%%%%%%%%%%%%%%%%%%%%%%%%%%%%%%%%%%%%%%%%%%%
%%
\begin{table}[t!]
\caption{Confusion matrices for Case 1 when $M = 3$.}
\vspace{2pt}
\label{tab:Case1_M_3}
\center
\scalebox{0.56}{
\begin{tabular}{|l|l|l|l|l|l|l|l|}
\hline
\multirow{3}{*}{\begin{tabular}[c]{@{}l@{}}LOS\\ using\\ SVM\end{tabular}} &
   &
  $\{v_1\}$ &
  $\{v_2, v_3, v_4, v_5\}$ &
  \multirow{3}{*}{\begin{tabular}[c]{@{}l@{}}NLOS\\ using\\ SVM\end{tabular}} &
   &
  $\{v_1\}$ &
  $\{v_2, v_3, v_4, v_5\}$ \\ \cline{2-4} \cline{6-8} 
 & $\{v_1\}$  & 0 & 5  &  & $\{v_1\}$  & 0 & 5  \\ \cline{2-4} \cline{6-8} 
 & $\{v_2, v_3, v_4, v_5\}$  & 0 & 13 &  & $\{v_2, v_3, v_4, v_5\}$  & 0 & 13 \\ \hline
\multirow{3}{*}{\begin{tabular}[c]{@{}l@{}}LOS\\ using\\ NN\end{tabular}} &
   &
  $\{v_1\}$ &
  $\{v_2, v_3, v_4, v_5\}$  &
  \multirow{3}{*}{\begin{tabular}[c]{@{}l@{}}NLOS\\ using\\ NN\end{tabular}} &
   &
  $\{v_1\}$ &
  $\{v_2, v_3, v_4, v_5\}$  \\ \cline{2-4} \cline{6-8} 
 & $\{v_1\}$  &  4 & 1  &  & $\{v_1\}$  & 4 & 1  \\ \cline{2-4} \cline{6-8} 
 & $\{v_2, v_3, v_4, v_5\}$  & 0 & 13 &  & $\{v_2, v_3, v_4, v_5\}$  & 2 & 11 \\ \hline
\end{tabular}}
\vspace*{-0.15in}
\end{table}
%%%
\begin{table}[t!]
\caption{Confusion matrices for Case 2 when $M = 3$.}
\vspace{2pt}
\label{tab:Case2_M_3}
\center
\scalebox{0.55}{
\begin{tabular}{|l|l|l|l|l|l|l|l|}
\hline
\multirow{3}{*}{\begin{tabular}[c]{@{}l@{}}LOS\\ using\\ SVM\end{tabular}} &
   &
  $\{v_3\}$ &
  $\{v_2\}$ &
  \multirow{3}{*}{\begin{tabular}[c]{@{}l@{}}NLOS\\ using\\ SVM\end{tabular}} &
   &
  $\{v_3\}$ &
  $\{v_2\}$ \\ \cline{2-4} \cline{6-8} 
 & $\{v_3\}$  & 8 & 0  &  & $\{v_3\}$  & 7 & 1  \\ \cline{2-4} \cline{6-8} 
 & $\{v_2\}$  & 1 & 2 &  & $\{v_2\}$  & 0 & 3 \\ \hline
\multirow{3}{*}{\begin{tabular}[c]{@{}l@{}}LOS\\ using\\ NN\end{tabular}} &
   &
  $\{v_3\}$ &
  $\{v_2\}$  &
  \multirow{3}{*}{\begin{tabular}[c]{@{}l@{}}NLOS\\ using\\ NN\end{tabular}} &
   &
  $\{v_3\}$ &
  $\{v_2\}$  \\ \cline{2-4} \cline{6-8} 
 & $\{v_3\}$  & 7 & 1  &  & $\{v_3\}$  & 8 & 0  \\ \cline{2-4} \cline{6-8} 
 & $\{v_2\}$  & 0 & 3 &  & $\{v_2\}$  & 0 & 3 \\ \hline
\end{tabular}}
\vspace*{-0.15in}
\end{table}
\begin{table}[t!]
\caption{Confusion matrices for Case 3 when $M = 3$.}
\vspace{2pt}
\label{tab:Case3_M_3}
\center
\scalebox{0.55}{
\begin{tabular}{|l|l|l|l|l|l|l|l|}
\hline
\multirow{3}{*}{\begin{tabular}[c]{@{}l@{}}LOS\\ using\\ SVM\end{tabular}} &
   &
  $\{v_3, v_4, v_5\}$ &
  $\{v_2\}$ &
  \multirow{3}{*}{\begin{tabular}[c]{@{}l@{}}NLOS\\ using\\ SVM\end{tabular}} &
   &
  $\{v_3, v_4, v_5\}$ &
  $\{v_2\}$ \\ \cline{2-4} \cline{6-8} 
 & $\{v_3, v_4, v_5\}$  & 11 & 0  &  & $\{v_3, v_4, v_5\}$  & 11 & 0  \\ \cline{2-4} \cline{6-8} 
 & $\{v_2\}$  & 2 & 2 &  & $\{v_2\}$  & 3 & 1 \\ \hline
\multirow{3}{*}{\begin{tabular}[c]{@{}l@{}}LOS\\ using\\ NN\end{tabular}} &
   &
  $\{v_3, v_4, v_5\}$ &
  $\{v_2\}$  &
  \multirow{3}{*}{\begin{tabular}[c]{@{}l@{}}NLOS\\ using\\ NN\end{tabular}} &
   &
  $\{v_3, v_4, v_5\}$ &
  $\{v_2\}$  \\ \cline{2-4} \cline{6-8} 
 & $\{v_3, v_4, v_5\}$  & 10 & 1  &  & $\{v_3, v_4, v_5\}$  & 11 & 0  \\ \cline{2-4} \cline{6-8} 
 & $\{v_2\}$  & 0 & 4 &  & $\{v_2\}$  & 1 & 3 \\ \hline
\end{tabular}}
\vspace*{-0.15in}
\end{table}
%%%%%%%%%%%%%%%%%%%%%%%%%%%%%%%%%%%%%%%%%%%%%%%%%
%\vspace*{-0.11in}
\section{Conclusion}
\vspace*{-0.13in}
We have presented machine-learning algorithms for the classification of human and non-human activities using wireless signals received at a massive MIMO base station.  We tested the methods on data obtained from a measurement campaign conducted indoors, using the $100$-antenna LuMaMi massive MIMO testbed operating at $3.7$ GHz carrier frequency \cite{lumami_journal}.  Experiments were conducted both in LOS and NLOS  scenarios as well as in both static and dynamic environments.  In the experimental tests, our proposed algorithms could successfully distinguish between human (e.g. a person dancing) and non-human (e.g. a spinning bike wheel) activities. Furthermore, the classification performance when using all $M=100$ antennas at the base station was significantly better compared to when using only $M=3$ antennas (a small subset of the array). This suggests that the spatial resolution capabilities offered by massive MIMO technology has the potential to significantly enhance the accuracy in wireless sensing applications. 
% -------------------------------------------------------------------------
\vfill
\pagebreak
%\pagebreak

%\section{REFERENCES}
%\label{sec:refs}

% References should be produced using the bibtex program from suitable
% BiBTeX files (here: strings, refs, manuals). The IEEEbib.bst bibliography
% style file from IEEE produces unsorted bibliography list.
% -------------------------------------------------------------------------

\bibliographystyle{IEEEbib}
\bibliography{icassp_liu-lu}

\end{document}